# Simulations of surface stress effects in nanoscale single crystals

*V. Zadin[a,b,1], M. Veske[a,b], S. Vigonski[a,b], V. Jansson[b], J. Muszinsky[b], S.Parviainen[b], A. Aabloo[a], F.Djurabekova[b]*

[a]Intelligent Materials and Systems Lab, Institute of Technology, University of Tartu, Nooruse 1, 50411 Tartu, Estonia
[b]Helsinki Institute of Physics and Department of Physics, P.O.Box 43 (Pietari Kalminkatu 2), FI-00014 University of Helsinki, Finland

**Abstract**

Onset of vacuum arcing near a metal surface is often associated with nanoscale asperities, which may dynamically appear due to different processes ongoing in the surface and subsurface layers in the presence of high electric fields. Thermally activated processes, as well as plastic deformation caused by tensile stress due to an applied electric field, are usually not accessible by atomistic simulations because of long time needed for these processes to occur. On the other hand, finite element methods, able to describe the process of plastic deformations in materials at realistic stresses, do not include surface properties. The latter are particularly important for the problems where the surface plays crucial role in the studied process, as for instance, in case of plastic deformations at a nanovoid. In the current study by means of molecular dynamics and finite element simulations we analyse the stress distribution in single crystal copper containing a nanovoid buried deep under the surface. We have developed a methodology to incorporate the surface effects into the solid mechanics framework by utilizing elastic properties of crystals, pre-calculated using molecular dynamic simulations. The method leads to computationally efficient stress calculations and can be easily implemented in commercially available finite element software, making it an attractive analysis tool.

---

[1]Corresponding author: vahur.zadin@ut.ee

# 1 Introduction

Interaction of an electric field with materials becomes increasingly important in a number of modern technologies under development. Some examples are the Compact Linear Collider (CLIC) [1,2], free electron lasers [3], fusion reactors [4,5] and atom probe tomography [6]. One frequently observed problem in these high electric field systems are vacuum arcs - the electric discharges in vacuum between two electrodes. These are also known as electrical breakdowns since they are commonly detected as a sudden voltage drop and a high electron current accompanied by significant power consumption. The electrical breakdowns lead to significant surface damage in high electric field devices and pose major limitations for the strength of the fields [7,8]. Measurements of electron field emission currents from macroscopically flat surfaces suggest that some surface irregularities with diameters in the range of 17-25 nm [9] and heights up to 100 nm [10,11] must be present to serve as field emitting tips. While such nanoscale emitters have never been observed experimentally [10], they are typically assumed to be possible sources of strong field emitting currents, precursors to a breakdown event. For example, Norem et al., used molecular dynamics (MD) to simulate possible breakdown mechanisms [12,13] and demonstrated in [12] evaporation of large clusters of atoms under high electric field while the finite element method (FEM) was utilized in [14,15] to investigate field enhancement effects due to micro cracks caused by fatigue in the material surface. These works provide important insight in possible breakdown initiation mechanisms but do not explain what triggers initial surface roughening. Several experimental and theoretical works link material structure and properties to the breakdown initiation mechanism - Descoeudres et al. [8] showed a correlation between the lattice type and its tolerance to the breakdowns; Nordlund and Djurabekova [16] linked breakdown probability to the dislocation motion in the material. Further MD studies by Pohjonen et al. demonstrated surface modification of Cu containing a subsurface nanovoid due to the effect of an electric field. [17,18]. The short simulation times in MD simulations required exaggeration of applied electric fields in order to observe any dislocation activity [17,18] These fields were possible to decrease considerably [19] by using FEM to simulate dynamic plastic deformations of Cu surface with an applied external electric field.

To overcome the vast differences in time and length scales between experiments and simulations, a multi-scale approach is needed. While experimental time expands to seconds or even minutes [10], the breakdown process itself is relatively fast: when initiated, the process culminates in a sub-microsecond time interval [20,21]. Methods with sufficient spatial resolution, such as MD, alone are still not sufficient to study the process due to the limitation of very short timescales – from pico- to nanoseconds. The time limitation can be overcome by using continuum methods, tailored to include nanoscale phenomena. For instance, it was shown [22] that these methods can be used for nanostructures of the size 2 -10 nm, if the effects of surface stress are resolved [18,22–24]. Nanoscale size effects have already been taken into account previously in different studies, in [25], the surface model was implemented using interatomic potential. In [23], a XFEM-based approach was used to simulate nanoscale size effects. In [26], an embedded atom hyper-elastic constitutive model was developed by He and Li. A different approach was used in [27] where the elastic properties of surfaces were calculated by MD methods. The onset of vacuum breakdown is however a complex phenomenon, which involves multiphysics processes, such as emission currents, Joule heating of material, and interaction of material surface with applied electric fields [28,29]. Having in mind the complexity of the problem, the existing approaches, although accurate and reliable, make the simulations computationally heavy.

While in [30] we implemented the surface stress model similarly to [31] and analyzed the sensitivity of different voids with uniform crystal faces, we will in this work propose an enhanced FEM bulk material model with an incorporated model of the surface stress in a fully coupled manner for the modelling of the mechanical behaviour of nanoscale FCC crystals. The model is capable of automatically approximating any crystal surface present in this model, as well as taking into account the size effects arising from nanoscale defects. As a suitable compromise we rely on MD simulations in order to obtain the elastic parameters for the surface region. This approach allows for simple implementations and low computational costs while still providing detailed information of the influence of the surface stress. We use a deep subsurface spherical void as a test stress concentrator and we compare the FEM model with MD simulations to analyse the stress distributions in nanoscale Cu.

## 2   Materials and methods

### 2.1   Molecular dynamics simulations

To study the material behaviour and possible protrusion formation mechanisms in a single Cu crystal, MD simulations were performed using the LAMMPS [32,33] classical molecular dynamics code. The atomic interactions were modelled using the embedded atom method (EAM) potential by Sabochick et al. [34]. Simulation results were visualized with the open-source OVITO software [35].

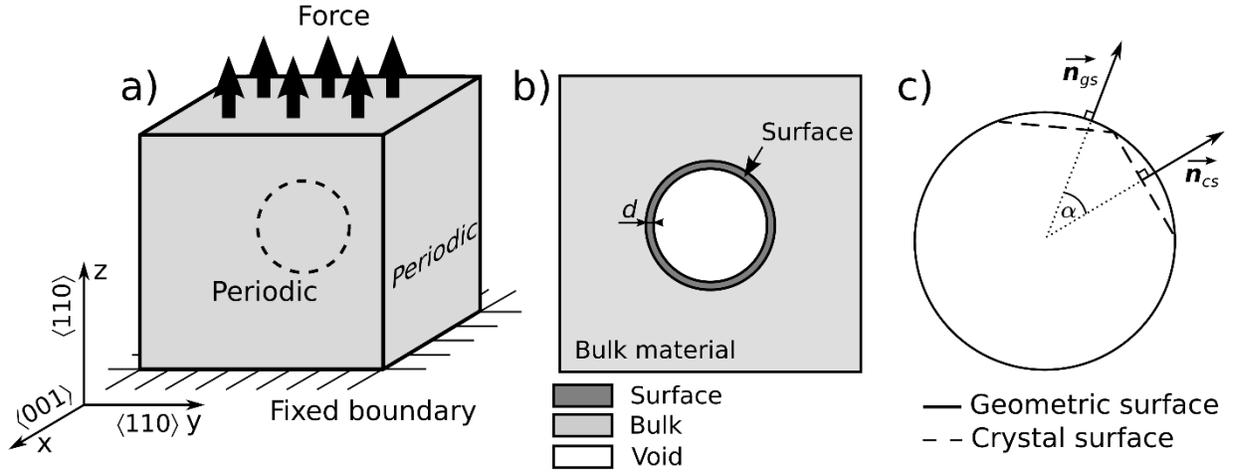

**Figure 1.** The simulation box set-up in MD and FEM simulations (a-b) and schematic representation of crystal face detection algorithm (c). Simulation cell is composed of FCC Cu single crystal. External tensile stress is applied to the top 2 layers of atoms, 3 bottom layers of atoms are fixed. The simulation cell is periodic in *x* and *y* directions.

We conducted MD simulations to investigate the effect of surface stress on the inner surface of a spherical void buried deep under the surface. The void was considered to be buried under the surface, if no interactions of the strain fields between the surfaces of the material and the void were observed. The crystallographic orientation of the Cu cell was chosen so that one of the {111} slip planes along the {110} direction would intersect the surface perpendicularly. The MD simulations were performed by using two simulation cells of dimensions 34.7x34.7x32.7 nm³ (3.3 million atoms) and 47x47x47 nm³ (8.8 million atoms) to investigate possible adverse effects arising from the use of periodic boundaries. For the smaller cell the periodic boundaries were used only in lateral directions, while the bottom of the cell was fixed and the top was treated as open surface, where the force imitating the Lorenz force of the applied electric field (see below) was exerted. For the bigger cell periodic boundary was applied in all directions. A 5 nm void was placed in the middle of each cell. The conducted MD simulations and their respective box sizes, void radii and boundary conditions are summarized in Table 1, geometry of the simulation set-up is presented in Figure 1.

The electric field and the electrons of the surface atoms interact due to the Lorenz force. The interaction strength is characterized by the Maxwell stress tensor, which, in case of DC electric fields, leads to the following expression for the force acting on the surface of the material [17,36]

$$\mathbf{F} = \frac{1}{2}\varepsilon_0 |\mathbf{E_0}|^2 A\mathbf{n} \tag{1}$$

where $\varepsilon_0$ is the vacuum permittivity, $\mathbf{E_0}$ is the electric field, $A$ surface area and $\mathbf{n}$ the surface normal vector.

| Box size | 34.7x34.7x32.7 nm$^3$ | 47x47x47nm |
|---|---|---|
| Number of atoms | 3.3 million atoms | 8.8 million atoms |
| Void radius | $r_{void} = 5$ nm | $r_{void} = 5$ nm |
| Boundary conditions | Periodic boundaries in $x$ and $y$ directions. Fixed bottom, Free/applied force in $z$ direction | All boundaries are periodic |
| Temperature | 293.15 K | 293.15 K |

Table 1 The details of MD simulations.

The simulations were conducted using 2 fs time step. In all cases the systems were equilibrated during 20 ps while the temperature was controlled by the Berendsen thermostat with a damping parameter of 0.1 ps. External stress was ramped during the 150 ps period by a gradual increase of the force acting on surface atoms until the final stress reached the value of 1.35 GPa. After that, the simulations were continued at the constant force for another 300 ps in order to obtain sufficient amount of data for the stress time average. Since the stress distribution was of particular interest, a lower temperature (293 K) and an external force was used to hinder the nucleation of dislocations that could cause the distortion of stress distribution.

## 2.2 Stress calculations in MD simulations

Different approaches to calculate the stress in the atomistic simulations exists. One of the most known methods is the virial theorem, reaching back to the work of Clausius and Maxwell, or more modern approaches, such as the ones developed by Hardy, Lutsko or Tsai [37–39]. While the latter ones can be considered more advanced and accurate than virial theorem, excellent agreement between Hardy and virial stress approaches has been demonstrated if sufficient spatial or temporal averaging is applied [40,41]. Thus, in current work we rely on the virial stress, due to the readily available implementations in many molecular dynamics codes. More specifically, in MD calculations conducted using Lammps, the symmetric 6 component stress tensor for atom $i$ is given by following equation [42,43]:

$$S_{ab} = -\left[ m v_a v_b + \frac{1}{2} \sum_{n=1}^{N_p} (r_{1a} F_{1b} + r_{2a} F_{2b}) + \frac{1}{2} \sum_{n=1}^{N_b} (r_{1a} F_{1b} + r_{2a} F_{2b}) + \sum_{n=1}^{N_f} r_{ia} F_{ib} \right] \tag{2}$$

where $a$ and $b$ assume corresponding $x$, $y$, $z$ values, the first term represents the contribution of kinetic energy from atom $i$, the second term the pairwise energy contribution. Finally, $N_p$ is the number of neighbours of atom $i$, $r_1$ and $r_2$ are the positions of atoms; $F_1$ and $F_2$ are the forces affecting the atoms due to pairwise interactions and internal constraints for atom $i$ are fixed by $N_f$.

The estimate for continuum mechanical (Cauchy) stress was calculated from the virials for each atom, obtained from the MD simulations and correspondingly normalized with homogenization volume, that in this study is considered to be equal to the average atomic volume [40,41]. The atomic volumes were obtained by finding non-overlaping Voronoi cells of atoms using the open-source Voro++ software [44]. The time averaging was conducted over 12000 and 15000 sequential time steps or 60 ps and 75 ps, respectively.

In addition to stress calculations, strain estimations are conducted according to [45,46] using OVITO software.

## 2.3 Stress calculations using finite element method

### 2.3.1 Bulk stress calculations

Successful simulation of surface stress requires combination of material bulk and surface models. In current approach, we assume linear elasticity and consider the bulk and surface models to be separate but coupled by boundary conditions. In bulk models large deformations are assumed. The material deformation is represented by the deformation gradient tensor ($F_{ij}$), connecting the deformed ($x_i$) and non-deformed configurations ($X_i$) [47,48]

$$x_i = F_{ij} X_j \tag{3}$$

The corresponding Green-Lagrange strain ($\varepsilon$) is calculated as

$$\varepsilon = \frac{1}{2} \left( F^T F - I \right) \tag{4}$$

Since large deformations are assumed, second Piola-Kirchoff stresses ($S$) are used [47,48]

$$S = C : \varepsilon_{el} \tag{5}$$

The second Piola-Kirchoff stress, Cauchy stress ($\sigma$) and first Piola-Kirchoff stress ($P$) are related as $S = F^{-1}P$ and $\sigma = J^{-1}PF^T = J^{-1}FSF^T$, where $J = \det(F) = V/V_0$ is the ratio between the deformed and non-deformed volumes.

Since the simulated material is single crystal copper, we use anisotropic elasticity. The elastic parameters of material parameters are specified in the Table 2 [47]

| $C_{11}$(GPa) | $C_{44}$(GPa) | $C_{12}$(GPa) | $d\alpha$(rad) | $h$ (rad) | $d_{ref}$(Å) |
|---|---|---|---|---|---|
| 168.4 | 75.4 | 121.4 | 0.24 | 0.3 | 3 |

**Table 2 The elastic properties of bulk material [47] and geometric parameters of the simulated surface layer in the calculations.**

The stress distribution in the material is calculated using the principle of a virtual work – the work from all external loads is equal to the virtual work from internal strains.

### 2.3.2 Surface stress

The surface stress and the surface energy are related as [27,49]

$$\tau(e) = \gamma + \frac{\partial \gamma}{\partial e}, \tag{6}$$

where $e$ is the elastic strain of the surface due to the surface tensile stress, $\gamma$ the surface energy and $\tau$ the surface stress. To simulate the influence of the surface stress on the material behaviour, the bulk model must be extended to describe the surface properties as well. When FEM is considered, this can be done by geometrically resolving the surface layer and specifying explicitly the initial surface stress at zero strain, as well as the surface elastic properties. However, when the surface layer is modelled geometrically, a large number of elements are required to handle the mesh generation and stress calculation in this area. Another possible approach is to model the material surface mathematically using the thin layer approximation (the thickness of the layer is much smaller than the rest of the surface) with shell or membrane elements [50]. This approach unavoidably builds in additional mathematical complexity, which may even introduce some numerical errors in the bulk model, however, it significantly reduces the requirements for the mesh density.

In current study, we handle the material surface by using mathematical modelling and the membrane elements. Thus, to simulate the surface layer behaviour, we add extra model available only at the geometrical surface of the material. The surface model setup is similar to the bulk model but has several important differences. Instead of simulating the area close to surface in fully 3D approach, we use the thin film formulation, where only tangential components of stress and strain over the surface change [50]. This formulation is realized using 3D plane stress elements, able to deform both in-plane and out-plane direction while the bending stiffness is neglected. It can be ignored in the surface model, as the surface processes cannot be treated independently but are tightly coupled to the bulk material behaviour (discussed below) [27]. As a result, in the current model, the surface layer bending behaviour is controlled by the bulk material.

The surface layer thickness ($d$) in the normal direction is considered to be small, giving constant stress and strain distributions in that direction. As in the bulk case, if the surface undergoes deformation, the relation between deformed ($x_i$) and non-deformed ($X_i$) configurations can be expressed using the deformation gradient ($F^s$), so that $x_i = F^s_{ij} X_j$ and the corresponding Green-Lagrange strain tensor can be calculated as:

$$e = \frac{1}{2}\left(F^{sT}F^s - \mathrm{I}\right)$$ (7)

Since the material surface undergoes deformation, the ratio of areas between deformed and non-deformed configurations is:

$$J = \sqrt{\det(F^T F)} = \sqrt{\det(I + 2e)}$$ (8)

Finally, the surface stress tensor is calculated as [27]

$$\tau_{ij} = \tau^0_{ij} + S_{ijkl}e_{kl},$$ (9)

where $\tau^0_{ij}$ is the initial surface stress and $S$ the fourth order surface elasticity tensor. Since the membrane approximation is used for the surface stress model, the thickness ($d$) of the membrane must be included in calculations.

The membrane thickness $d$ is also used to introduce the size dependence, characteristic to the nanoscale systems, into the model (and avoid scale invariance during upscaling of the geometry). We consider $d = d_{ref} \cdot \dfrac{\kappa}{\kappa_{ref}}$ , where $\kappa$ is the curvature of the void and the subscript *ref* characterizes the reference configuration – a void with 5 nm radius. The interface thickness $d$ was initially obtained through the comparison with MD simulations.

The incorporation of the finite size effects was later tested by comparing the simulation results with the analytical model, presented in [51]. The comparison was obtained by conducting series of simulations using different external stresses (3.05 GPa and 4.58 GPa), while the depth of the void was increased (starting from the surface with 0.2 nm steps) and the radius of the void was scaled from 4 nm to 20 nm with 1nm steps.

### 2.3.3  Coupling the surface and bulk stress models

To obtain the final stress-strain distribution in the nanoscale material due to the surface stress influence, both surface and bulk models must be coupled. The coupling is achieved by binding the two stress models using the boundary conditions and solving a nonlinear system of equations. As the deformation of the bulk material also causes the deformation of the surface, the bulk deformation is carried over to the surface model - every point in the surface model undergoes the deformation experienced by the boundary of the bulk model, leading to the predefined displacement in the surface stress model:

$$\mathbf{u}_{surf}(x,y,z) = \mathbf{u}_{bulk}(x,y,z) \tag{10}$$

The deformation of the surface leads to accumulation of the surface stress according to eq. 9. The achieved surface stress is coupled to the bulk model by incorporating it to the pressure load boundary condition. The resulting forces at the boundary of the bulk model distribute as:

$$\mathbf{F} = p\mathbf{n}J \ , \tag{11}$$

where $p$ is the pressure calculated in the surface stress model, $\mathbf{n}$ the unit normal vector in the deformed configuration and $J$ defined in eq. 8. The surface stress boundary condition is implemented as a follower load, meaning that the changes of the surface areas during the calculations are monitored and the loads are applied in the deformed frame (not in the initial, non-deformed one).

### 2.3.4  Crystal face detection

Important part of the model implementation is the identification of crystal faces constituting the surface, since both the initial surface stress and the elastic properties of the surface are crystal orientation dependent. Two possible strategies can be used for that. The first one relies on the geometry setup – the initial geometry is well defined, where all the boundaries correspond directly to the crystal faces. In this case, all the surface properties are pre-defined by the user. The shortcoming of this method is the need to manually specify the crystal faces by constructing sufficiently accurate geometry.

In the second case, the geometry doesn't have boundaries corresponding directly to the crystal faces, but the boundaries follow the actual crystal faces only approximately. For example, if a perfectly spherical void is considered in the finite element model, this is an approximation of the actual material defect. In an actual crystal structure, a perfectly spherical hole cannot exist due to the atomistic nature of the matter. Instead, the surface of the defect (void) consists of patches of differently oriented crystal faces. In this case, a special algorithm must be used first to determine the crystal faces constituting the surface and secondly translate this information to construct the surface model which includes the corresponding effective surface elastic properties.

In current work we use the second approach – we reconstruct the crystal surface approximately using the identification algorithm presented below. By choosing this approach, we gain computational robustness and flexibility on expense of some accuracy.

| Crystal planes ($k$) | Nonzero components of initial surface stress tensor $\tau_{11}^0$ (eV/ Å$^2$), $\tau_{22}^0$ (eV/ Å$^2$) | Bulk modulus $S_{iijj}$ (eV/ Å$^2$) | Shear modulus $S_{1212}$ (eV/ Å$^2$) |
|---|---|---|---|
| {001} | 0.0649, 0.0649 | 0.017 | -0.063 |
| {111} | 0.0343, 0.0343 | -0.527 | 0.009 |
| {110} | 0.0621, 0.0373 | -2.428 | -0.352 |
| Rest of the surface - {112} | 0.0474, 0.0382 | -1.641 | -0.252 |

**Table 3 Elastic parameters of crystal surfaces in the FEM simulations [27]**

The crystal face identification algorithm utilizes the Miller indices, characterizing every crystal plane. The indices provide the normal vectors to the crystal planes on surface of the studied void. In a FCC single crystal, we identify the crystal planes shown in Table 3, which are presently included in the proposed algorithm.

In the following, we refer to the *geometrical surface* as the surface modelled by using conventional modelling tools available in the used FEM software. The *crystal surface,* however, is the surface which is constructed according to the proposed algorithm with the surface properties simulated by MD in the presence of a volumetric defect. In case of the geometrical surface, we do not model the crystal faces directly, while in case of the crystal surface, the faces appear naturally. This is illustrated in Figure 1c, where solid line represents the geometry surface and dashed line the actual crystal surface.

To identify approximately the crystal faces on the geometrical surface, we consider the surface normal vectors of the geometrical model $\mathbf{n}_{gs}$ and the normal vector $\mathbf{n}_{cs}^{k}$ of the crystal plane $k;$ the latter arising from the orientation of material. In the case of parallel normal vectors, the geometrical model represents the crystal surface accurately; otherwise approximately. To reconstruct a valid model of the crystal surface and identify the actual crystal faces $k$ on the geometrical surface, we first find the angles between $\mathbf{n}_{gs}$ and $\mathbf{n}_{cs}^{k}$ (see Figure 1b)

$$\alpha^{k} = \cos^{-1}\left(\frac{\left\langle \mathbf{n}_{cs}^{k}, \mathbf{n}_{gs}\right\rangle}{\left|\mathbf{n}_{cs}^{k}\right| \cdot \left|\mathbf{n}_{gs}\right|}\right) \tag{12}$$

Both, $\mathbf{n}_{gs}$ and $\mathbf{n}_{cs}^{k}$ must be specified in the material coordinate system (laboratory coordinate system), so that they follow the correct crystal orientation in the geometry. Finally we limit the maximum angle between $\mathbf{n}_{gs}$ and $\mathbf{n}_{fs}^{k}$ for that part of the geometrical surface to find the crystal face $k$. To do it, we use the smoothed Heaviside function (due to the practical considerations of avoiding sharp transitions, possibly leading to numerical problems). The smoothed Heaviside function, *flc2hs*, in Comsol Multiphysics is defined as:

$$H(\phi, h) = flc2hs(\phi, h) = \left(\frac{\phi}{h} > -1\right)\left(\frac{\phi}{h} < 1\right)\left(\frac{1}{2} + \frac{\phi}{h}\left(\frac{15}{16} - \frac{5}{8}\left(\frac{\phi}{h}\right)^{2} + \frac{3}{16}\left(\frac{\phi}{h}\right)^{4}\right)\right) + \left(\frac{\phi}{h} \geq 1\right) \tag{13}$$

where $h$ is the width of the transition region from zero to one. Now, the crystal faces are identified according to:

$$\Phi^{k} = H\left(\left|\alpha^{k}\right| - d\alpha, h\right) \tag{14}$$

Where $d\alpha$ is the maximum allowed angle between the normal vectors of the geometrical and the crystal surfaces. Finally, all crystal faces in a given crystal plane family are collected by summing the $\Phi_i$. For example, for {100} plane family:

$$\Phi^{[100]} = \Phi^{(100)} + \Phi^{(010)} + \Phi^{(001)} \tag{15}$$

Since the Heaviside function is used to detect the crystal faces, value of $\Phi^k$ stays always between 0 and 1, with 1 corresponding to the crystal plane $k$. Thus, the effective elastic properties of the surfaces, reflecting the collective behaviour of all present crystal faces, are now easily expressed as nonlinear functions of identified crystal faces:

$$\tau_{11}^* = \sum_q \tau_{11}^q \Phi^q \ , \quad \tau_{22}^* = \sum_q \tau_{22}^q \Phi^q \ \text{ and } \ S_{ijlm}^* = \sum_q S_{ijlm}^q \Phi^q \ , \tag{16}$$

where $q$ is the crystal plane family.

### 2.3.5 Finite element simulations

The finite element simulations were conducted using Comsol Multiphysics 4.4 and its Structural Mechanics and Membranes toolboxes [52]. Different meshes were tested during the calculations, and the final one consisted of ~132 000 mixed tetrahedral and hexahedral elements, with a minimum size of 0.5 nm and and a maximum size of 1.93 nm. The hexahedral elements were generated concentrically around the void until the sides of the simulation box were reached using swept meshing. The tetrahedral elements were used only far away from the void, in low stress and strain regions. The equations for the bulk material were solved using linear elements and for the surface behaviour, using quadratic elements. Since the obtained system of equations, describing the stress-strain relations in coupled bulk and surface models is highly nonlinear, the damped Newton type nonlinear solver [53] was used in conjunction with the Pardiso solver.

## 3    Results and discussion

### 3.1   Defect caused stress distribution and influence of the surface stress

To verify the calculations of FEM simulations, the shear stress obtained using FEM was compared to the results found from corresponding MD simulations. Both qualitative and quantitative comparisons of the results were conducted to evaluate the accuracy of the FEM model. The stress distributions are presented as cut planes from the simulated geometry; all slices are selected from the centre of the box and correspond to different crystallographic directions, specified in the figures. Both the MD and FEM simulations are coloured according to the shear stress component and are presented in Figures 2-6.

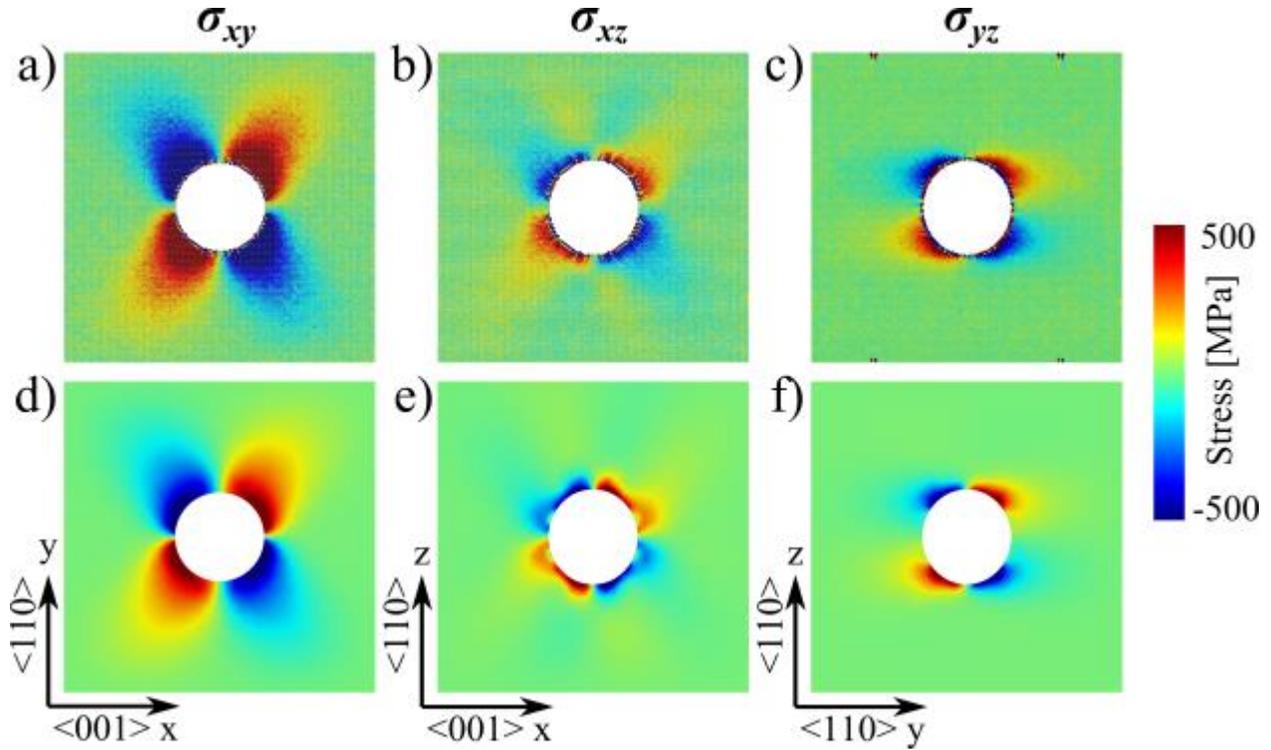

**Figure 2. Shear stress distribution around the void without external tensile stress. Top row – MD results; bottom row – FEM results. The relaxed material configurations demonstrate good agreement between the methods.**

The stress distributions without external force in case of a spherical void are presented in Figure 2. The presented shear stress distributions overlap well for both MD and FEM simulations for all presented stress tensor components, demonstrating qualitatively similar behaviour due to the influence of surface stress. The most significant difference between the MD and FEM simulations can be observed for the $\sigma_{xz}$ components near the leftmost and rightmost edges of void cross section– the crystal face interpolation algorithm causes slight distortion of material properties near the void surface leading to the deviation of stress distributions. However, the difference between the MD and FEM cases is small and the effect to the overall stress distribution caused by the void can be neglected.

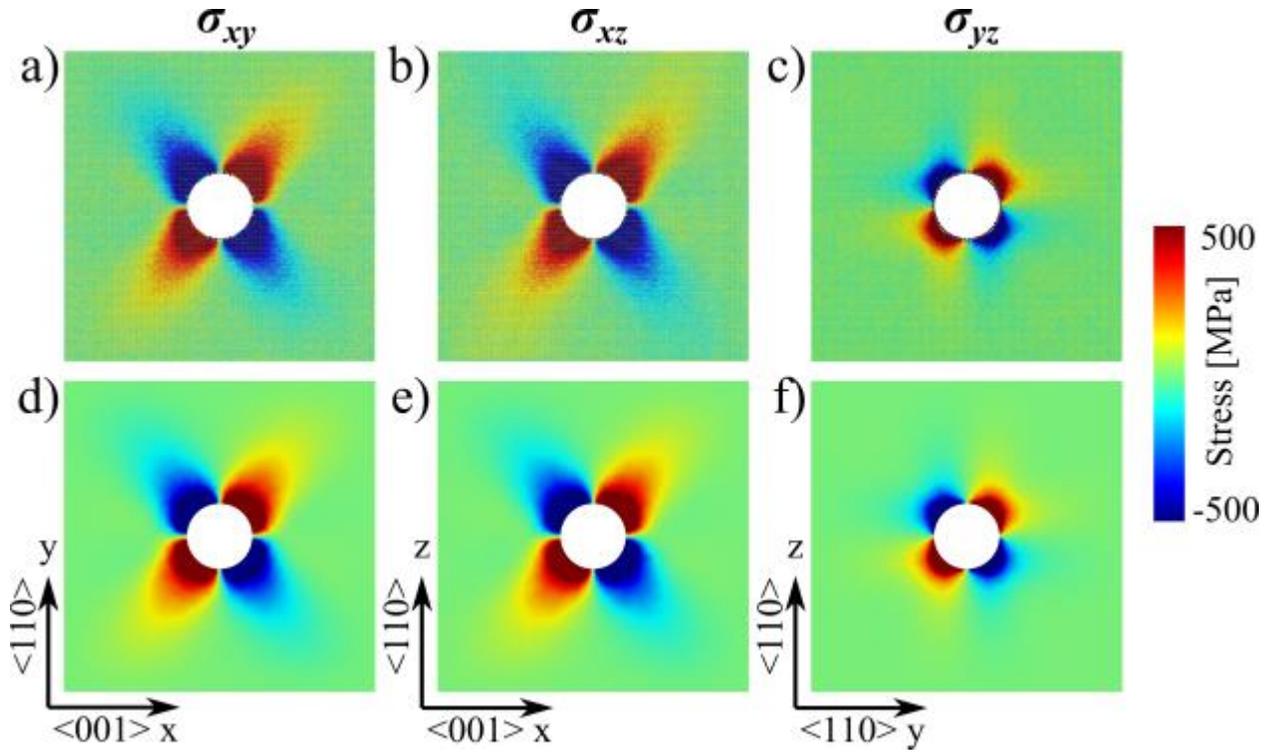

**Figure 3. Shear stress distribution in Cartesian slices around the void without external tensile stress in case of fully periodic boundary conditions. This configuration corresponds to void in bulk material and extends the results for arbitrary crystal orientation. Left column – σ$_{xy}$, middle column - σ$_{xz}$, right column – σ$_{yz}$. Upper row – MD results, lower row – FEM results.**

Although, in the previous studies we have considered a near surface void, the presence of the surface may influence the surface stress distribution near the void itself. Moreover, the crystal orientation with respect to the surface starts playing a critical role due to a high anisotropy of elastic properties of copper. To separate the internal surface stress distribution from the surface, as well as to avoid the necessity of rotation of the simulation cell in order to obtain the orientation independent result, we performed the current calculations of a void placed deep in the bulk. The results are presented in Figure 3. Again, the comparison of MD and FEM simulations demonstrates good qualitative agreement – all presented stress components behave similarly. Compared to Figure 2, only the σ$_{xy}$ behaviour is the same, as in both cases the material is already periodic in that direction. Other presented stress tensor components, σ$_{xz}$ and σ$_{yz}$, show considerable difference – indicating that limited material thickness around the void has significant influence on the stress distribution. However, changes in both MD and FEM agree, indicating that FEM can accurately capture the material behaviour in every crystal orientation.

To test the agreement between the methods for random direction, we conducted similar comparisons as in Fig. 3 for stress tensor components in the *yz* cut plane, rotated 45 degrees around *z* axis. Although not shown here, the stress distributions obtained by both methods are very similar consistently with the previous figures.

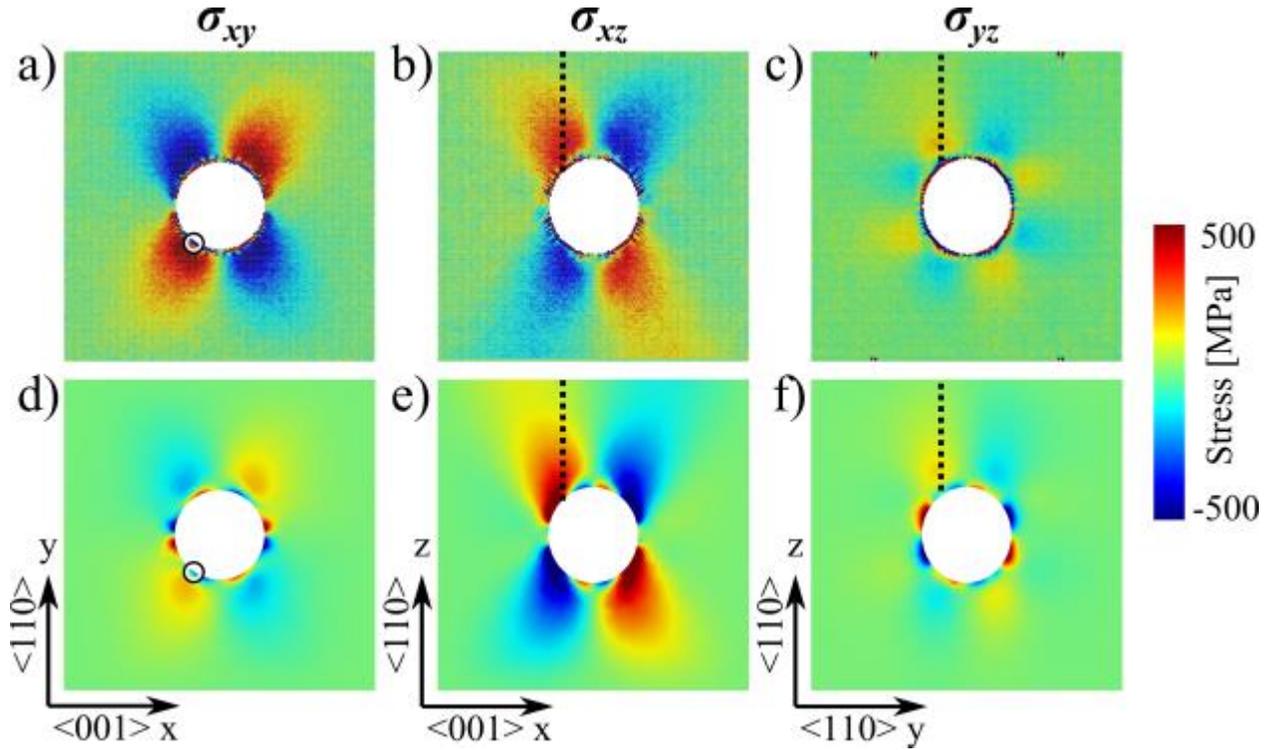

**Figure 4. Shear stress distribution around the void with external 1.35 GPa tensile stress. Left column – σ$_{xy}$, middle column - σ$_{xz}$, right column – σ$_{yz}$. Upper row – MD results, lower row – FEM results. The circular marker and dashed lines in the figures represent the data plotted in Fig 5.**

The stress distribution due to the combined effect of surface stress and external force are presented in Figure 4. Compared to the case without external force, the stress is distributed along the diagonal of the simulation cell and slightly stretched in the z direction. The main difference between the stress distributions can be explained by the approximations of the surface layers in FEM simulations. Compared to the case without an external stress (Figure 2), several short range distortions are visible in all presented stress components. These artefacts are influenced by the mesh used in the FEM calculations in combination with the crystal face identification algorithm as both of them introduce smoothing and approximations of surface properties into the model. While the surface mesh over the void has relatively small elements (0.5 nm), element size starts to grow in the bulk material and amplifies the effect surface artefacts deeper in the bulk. On the other hand, the MD stresses near the surface are influenced by the numerical artefacts as well. The MD stresses depend on the volumes associated with atoms, that can be estimated using the Voronoi cells. The latter are poorly defined for the surface atoms as the cells expand deep into the void. However, the influence of these artefacts to the overall stress distribution is small, even if they introduce additional numerical noise in the system. They can be viewed as acceptable trade-offs for keeping the size of the elements in the bulk material relatively large while guaranteeing fast and computationally efficient calculations.

## 3.2  Quantitative comparison of MD and FEM stress estimations

Finally, quantitative comparisons of the stress tensor components in MD and FEM simulations are presented in Figure 5. The data for the comparison was obtained by plotting the distribution of the shear stress component along the line in the z direction between the void surface (at the point where the shear stress has its maximum value) and the surface of the simulation cell. The MD results in Figure 5 are presented by blue lines and the FEM results by red lines (dashed without external force, solid with external force). The horizontal axis represents the distance from the void surface in ångström and the vertical axis gives the respective shear stress component $\sigma_{xy}$, $\sigma_{xz}$ or $\sigma_{yz}$ in GPa.

FEM and MD show again the same general trend in the stress distribution as in previous snapshots, demonstrating generally good agreement between the methods. All the presented lines can be divided into three main regions – the surface layer (<0.5 nm), close to the void region (distances under 2 nm) and the distant regions, further away. The maximum stress values stay always in the surface layer in both methods. However, since the both approaches need significant approximations in the surface layer area – MD due to the ill-defined atomic volumes for surface atoms and FEM due to uncertainties in choosing the surface layer thickness – the results differ by the factor of two.

When we consider the close to the void region, both models follow the same trends, while FEM generally underestimates the stresses. This can be caused by three mayor factors – the geometrical differences between the actual surfaces in MD and FEM, the accuracy of crystal surface reconstruction in FEM (we use only 4 crystal surface families) and the smoothing effects of the mesh. Since the mesh elements grow larger further away from the void, the smoothing occurs naturally as the stress calculated by FEM is the average over several atoms, while MD always provides the atomic stresses. Finally, the stress distribution is also affected by the elastic properties of the crystal surface.

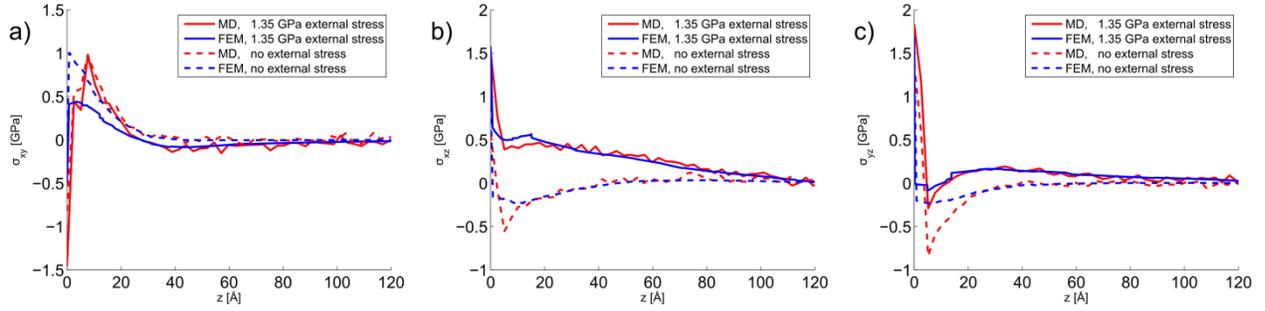

**Figure 5. Shear stress along the line from FEM and MD simulations. In x axis the distance from void surface in Å, in y axis a) shear stress $\sigma_{xy}$, b) shear stress $\sigma_{xz}$, c) shear stress $\sigma_{yz}$ in GPa. The data is plotted along the dashed line in z direction presented in Fig. 4 for corresponding stress components. $\sigma_{xy}$, is presented starting from circular marker in Figs. 4a and 4d in z direction (perpendicular to the schematics).**

Finally, the third distant region demonstrates excellent agreement between the MD and FEM simulations as all the stress values coincide.

Combination of data presented in Figure 2-5 demonstrates that the proposed modelling approach can effectively capture the influence of the void and its surface on the stress distribution in the bulk material. The computational time required for the FEM simulations can be very short - minutes (on dual core desktop PC, in case of coarser meshes), compared to hours of extensive parallel MD calculations. The FEM model augmented by the proposed surface model can be used to assess the ability of nanodefects present in real materials to concentrate the stress. It will also allow to estimate the possible weak points in the structure where the crystal lattice may yield a dislocation at the stresses well below the tensile strength of the material. However, the more rigorous simulations such as MD with appropriate interatomic potentials will be needed to simulate the mechanism of nucleation of dislocations or their reactions in detail.

Using FEM as a tool for the stress calculations also provides significant advantage compare to MD methods by the way the methodology is implemented. For instance, the stress calculated in MD per atom does not result in a single interpretation of the obtained data due to the ambiguity of definition of the cross sectional surface area of an atom. The finite element based approach, however, relies on a continuum approximation, eliminating completely the need to consider volumes or cross sectional areas of the atoms.

### 3.3 Estimation of strain in MD and FEM calculations

Finally, in addition to the stress estimates, strain obtained from both MD and FEM calculations are compared as well in Figure 6. There, subfigure a) represents the $\varepsilon_{yz}$ component of the strain tensor in MD simulations while Figure 6 b shows the same quantity in FEM calculations. Subfigure c) provides detailed quantitative comparison between FEM (blue line) and MD (red line) results for both $\varepsilon_{yz}$ (solid lines) and $\varepsilon_{xz}$ (dashed lines) at the dashed line presented in Figure 6a. Strain, as defined by Eqs. 3-4, relies only on the displacement and deformation gradient of the atoms, thus eliminating the need to use homogenization volumes as during the estimation of the stress.

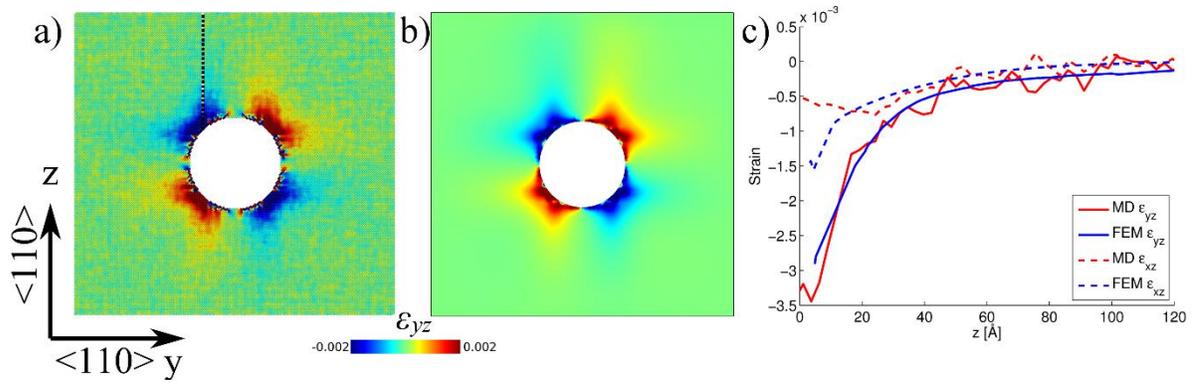

**Figure 6 Distribution of $\varepsilon_{yz}$ component of the strain tensor in MD (a) and FEM (b) simulations in case of fully periodic boundary conditions. The dashed line in subfigure a) shows the location of plotted data in c).**

However, due to the thermal movement of atoms at nonzero temperatures, time averaging is still needed. Similarly to the stress calculations, very good qualitative and quantitative agreement of the results is obtained with the results coinciding almost completely, in range of the nearest neighbor interpolation fluctuations of of MD strain components and providing final validation to the model.

### 3.4 Finite size effect in FEM models

To test the proposed model for ability to capture the finite size effect, we performed additional simulations of a near surface void with different radii in a single crystal Cu held under tensile stress. Stress was exerted on the Cu surface. The geometry and simulation condition were chosen to replicate those in MD simulations [40], where the analytical model to describe the dependence of void aspect ratio on the radius of the void was proposed. By the void ratio the ratio of the void radius $r$ to the void depth $h_{cyl}$, at which the maximal shear stress $\sigma_{zx}$ was sufficient to nucleate a dislocation at the void surface (for details see [40]). $h_{cyl}$ stands for the height of the cylinder above the void up to the surface of the material with the base circumscribed by the region of the maximal shear stress on the surface of the void. By plotting this parameter ($h_{cyl}/r$) against the radius of the void, $r$, it was demonstrated that at some size of the void, the void ratio $h_{cyl}/r$ became independent of the size of the void and the geometry became size invariant. Due to the size and time limitations, the MD results could not reproduce the entire curve $h_{cyl}/r(r)$. By developing the present model we have an opportunity to verify the finite size effect predicted by analytical model in [40]. If the curve can be reproduced by the present simulations, then the saturation of the dependence $h_{cyl}/r(r)$ can be explained only by the surface stress on the surface of the void, since this effect becomes negligible with the growth of a nanovoid.

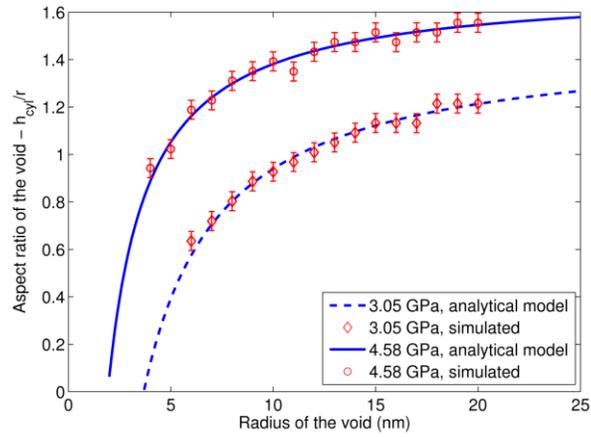

**Figure 7 Aspect ratio of a void needed for dislocation nucleation in FEM simulations vs. estimations by the analytical model published in [51]. Red circles and diamonds correspond to simulations; error bars represent the uncertainty due to the void depth increments (+/- 0.2 nm).**

The results of size dependence of the surface stress model and the comparison with the analytical predictions is presented in Figure 7 where the red markers represent the simulated data and blue lines represent the analytical model from [51], $h_{cyl}/r = a/\sqrt{2} + c/r$, where coefficients $a$ and $c$ are obtained from the simulation data. The error bars correspond to the void depth increments (0.2 nm) in the present calculations. The analysis was performed for two different external stress cases – 3.05 GPa and 4.58 GPa – with corresponding threshold stresses for 2 GPa and 3 GPa (corresponding to the increase of applied stress). In both cases, we see how the surface stress model follows the same behaviour as predicted by the analytical model very closely. While the FEM model succeeds in capturing accurately the behaviour of the size effects, the homogenization of the continuum approach leads to underestimation of stresses on the void surface and, consequently, to underestimation of the void aspect ratio as well. However, the close comparison between the analytical prediction [40] and the present FEM results support the analytical model and explain the size effects mostly due to the effects of surface stress on the surface of the void.

## 4   Conclusions

In this work, we have successfully developed a methodology to incorporate the surface stress effects into the solid mechanics framework. The developed methodology was tested by applying a tensile stress on a single Cu crystal containing a test stress concentrator, in the shape of a spherical void. The proposed approach utilizes pre-calculated elastic properties of crystal surfaces from the MD simulations as input parameters and allows calculating resulting stress distributions in a computationally efficient way. The comparison of stress estimates obtained using MD and FEM shows good qualitative and quantitative agreement between the two methods for all test cases, while additional validation is provided by the demonstration of similar agreement in the calculated shear strains as well. We have also demonstrated that the size effect, characteristic for nanoscale systems, is consistent with previous MD works. The methodology can be easily implemented using commercially available finite element solutions, making it an attractive analysis tool. Moreover, the needed input parameters for the elastic properties of the surfaces can be calculated using MD simulations or obtained from already published data. All these features make the proposed approach an attractive analysis tool for studying mechanical interactions in nanoscale materials.

## 5   Acknowledgments

This work is supported by Estonian Research Council grants PUT 57 and PUT 1372, and the national scholarship program Kristjan Jaak (funded and managed by Archimedes Foundation in collaboration with the Ministry of Education and Research of Estonia). Computing resources were provided by the Finnish IT Center for Science (CSC) and High Performance Computing Centre of University of Tartu. V. Jansson was supported by Academy of Finland (Grant No. 285382).